\documentclass[aps,preprint,twoside,nobalancelastpage,12pt,a4paper,onecolumn,secnumarabic]{revtex4}

\usepackage{amsmath}
\usepackage{amssymb}
\usepackage{amsfonts}
\usepackage{graphicx}

\begin{document}
\title[Multi-ion magnetoplasmas]{Multiscale structures in three species
magnetoplasmas with two positive ions}
\author{Shafa Ullah$^{\ast }$}
\author{Usman Shazad}
\author{M. Iqbal}
\affiliation{Department of Physics, University of Engineering and Technology, Lahore
54890, Pakistan}
\email{gulmiriwala@gmail.com}

\begin{abstract}
The self-organization in a multi-ion plasma composed of electrons and two
species of positively charged ions is investigated. It is shown that when
canonical vorticities and velocities of all the plasma fluids are aligned,
the magnetic field self-organizes to Quadruple Beltrami state (superposition
of four Beltrami fields). The self-organized magnetic and velocity fields
strongly depend on the relative strengths of the generalized vorticities,
flows, inertia and densities of the plasma species. Thus, it is possible to
generate a wide variety of multiscale magnetic field and flow structures. It
is also shown that relaxed magnetic fields and velocities can vary on vastly
different length scales simultaneously and are coupled together through
singular perturbation generated by Hall effect. In this multi Beltrami
self-organized states, then, the dynamo mechanism emerges naturally. The
scale separation also suggests the heating of the plasma through a
dissipative process. The work could be useful to study the dynamics and
morphology of the multiscale magnetic field configurations in laboratory and
astrophysical plasmas.
\end{abstract}

\maketitle

\section{Introduction}

The plasmas self-organize or relax to the lowest energy state under the
topological constraints of magnetic fields \cite%
{Hasegawa1986,Ortolani1993,Brown1997}. The relaxed state of a single fluid
plasma is called the Beltrami state and can be expressed mathematically in
terms of the equation $\mathbf{\nabla }\times \mathbf{B}=\lambda \mathbf{B}$%
, where $\mathbf{B}$ is called the Beltrami magnetic field. Here, $\lambda $
is a scalar field that must satisfy $\mathbf{B\cdot \nabla }\lambda $ to
insure $\mathbf{\nabla \cdot B}=0$. The Beltrami magnetic field represents a
stationary (flowless) force-free (Lorentz force vanishes) equilibrium state
of a single fluid plasma. Woltjer \cite{Woltjer1958} and Taylor \cite%
{Taylor1974} derived the Beltrami state by minimizing the magnetic energy
with the constraint that magnetic helicity is conserved. The Beltrami fields
are special type of fields and have been found very useful to study the
twisted magnetic fields observed in many astrophysical phenomena, for
instance, the magnetic flux ropes \cite{Saunders1984} and the galactic jets 
\cite{Konigl1985}. The Beltrami fields are used to describe circularly
polarized waves \cite{Chu1982}, to design the superconducting magnet \cite%
{Salingaros1990} and to express the equilibrium state of a magnetized
turbulent plasma \cite{Taylor1986}.

The single fluid relaxation theory was extended to two fluid plasmas \cite%
{Steinhauer1997,Steinhauer1998,Mahajan1998} to obtain the relaxed states
exhibiting strong flow and pressure gradients. The self-organized state of
two fluid (ions with inertialess electrons) plasmas is a non-force-free
state which can be written as a linear sum of two distinct Beltrami states
called Double Beltrami state (DB). Using variational principle, DB state can
be derived by minimizing the total energy subject to the constraints that
two magnetic helicities are conserved \cite%
{Steinhauer1997,Mahajan1998,Yoshida99,ZYoshida2001}. The DB state is
characterized by two scale parameters and appears as a result of strong
coupling between the magnetic fields and velocities. It ensures the
confinement of plasma through Bernoulli mechanism. The DB states offer a
much wider class of steady state equilibrium solutions which may help to
investigate and understand a variety of problems. The DB states have been
used to study the high $\beta $ (ratio of the kinetic to magnetic pressure)
equilibrium in tokamaks \cite{Yoshida2001} and H-modes in plasmas \cite%
{Guzdar2005}. The DB states have also been used in modelling the
astrophysical phenomena, for instance, solar arcade structures and loops,
eruptive events and solar flares \cite%
{Ohsaki2001,Ohsaki2002,Mahajan2002,Kagan2010}, heating of coronal plasmas 
\cite{Mahajan2001,Browning2003}, scale hierarchies in flows \cite%
{Yoshida2004}, dynamo mechanism \cite%
{Mininni2002,Mininni2003,Mahajanshata,Lingam2015} and turbulence \cite%
{Krishan2004}.

The inertia of plasma species plays an important role in the process of
self-organization. It has been shown that when inertial effects of both the
species in a two fluid plasma are taken into account, the Beltrami states
are increased in number. For example, in an electron-positron plasma, the
self-organized state is found to be composed of three distinct Beltrami
fields. The sum of three Beltrami fields is called Triple Beltrami (TB)
field \cite{Iqbal2008}. The diverse applications of Beltrami fields have
motivated the scientific community to study and analyze the relaxed
equilibria in plasmas containing more than two species. In three fluid
plasmas, when the inertia of all the plasma components is considered in the
dynamics, the self-organized state appears to be a Quadruple Beltrami state
(QB), which is a combination of four Beltrami states \cite%
{Shatashvili2016a,Gondal2017,Shatashvili2019a,Gondal2020,Gondal2020a}. The
multi-Beltrami relaxed states emerge naturally in a multi-fluids system. The
interesting features and implications of multi-Beltrami fields are discussed
in the Ref. \cite{Mahajan2015a}.

The present work is the extension of our previous work \cite{Iqbal2012} and
focuses on multi-ion three species plasma consisting of electrons with two
species of ions. Both species of the ions are positively charged and carry
different masses. Such type of multi-ion plasmas exist in different
planetary magnetospheres such as Saturn's magnetosphere \cite%
{Shemansky1992,Ip2000}, earth ionosphere and magnetosphere \cite%
{Zong2012,Nilsson2016,Luo2017}, solar wind \cite{Tracy2015}, polar solar
corona \cite{Zimbardo2011}, and in Mars environment \cite{Yamauchi2015}. For
industrial applications, multi-ion plasmas are created in magnetron
sputtering. The multi-ion plasmas created in magnetron sputtering are used
in plasma cleaning, etching, and thin-film deposition \cite%
{Tsuji1991,Budtz1999,Popescu2000,Popescu2000a,Budtz2001}.

The present work shows the possibility of creating multiscale self-organized
magnetic and velocity fields in the multi-ion (two positive species of ions
and electrons) plasmas by varying the mass and number densities of the two
group of positive ions in addition to their generalized magnetic fields and
flows. The system self-organizes to QB state when all the inertial and
non-inertial forces take part in the dynamics and can be transformed one
Beltrami state to another one on varying the fluid velocities, generalized
magnetic fields and densities of plasma components. It is also shown that
dynamo mechanism emerges naturally when the velocity and magnetic fields
varying at vastly different scale lengths couple each other. The scale
separation also suggests that plasma can be heated through a dissipative
mechanism when a relatively smooth magnetic fields are present.

The outline of the paper is as follows. In Sec. II, starting with the
macroscopic evolution equations of the plasma species, the steady state
solutions (Beltrami conditions) are derived for each plasma fluid. Using
Beltrami conditions and Ampere's Law, the QB state is derived and the
expressions for the four scale parameters (eigenvalues) are evaluated. In
Sec. III, the impact of Beltrami parameters on the formation of different
Beltrami states is presented. The analytical solution of the QB state for
the slab plasma is expressed in Sec. IV. The scale separation and its
implications are discussed in Sec. V. The summary of the work is presented
in Sec. VI.

\section{Model Equations}

Let us consider an incompressible magnetized multi-fluid plasma consisting
of electrons ($e$), lighter ions ($l$), and heavy ions ($h$). The number
densities of electrons, lighter ions, and heavy ions are $n_{e}$, $n_{l}$,
and $n_{h}$ respectively. The plasma is taken to be quasineutral and
satisfies%
\begin{equation}
1=Z_{h}N_{h}+Z_{l}N_{l},  \label{qn}
\end{equation}%
where $N_{h}=n_{h}/n_{e}$ and $N_{l}=n_{l}/n_{e}.$ The amount of charges
carried by lighter and heavier ions are represented by $Z_{l}$ and $Z_{h}$
respectively. The densities of plasma species are supposed to be constant.
The normalized momentum balance equations for the electrons, lighter ions
and heavier ions can be written as follows%
\begin{eqnarray}
\frac{\partial \mathbf{P}_{e}}{\partial t} &=&\mathbf{V}_{e}\times \left( 
\mathbf{\nabla }\times \mathbf{P}_{e}\right) -\mathbf{\nabla }\psi _{e},
\label{me} \\
\frac{\partial \mathbf{P}_{l}}{\partial t} &=&\mathbf{V}_{l}\times \left( 
\mathbf{\nabla }\times \mathbf{P}_{l}\right) -\mathbf{\nabla }\psi _{l},
\label{ml} \\
\frac{\partial \mathbf{P}_{h}}{\partial t} &=&\mathbf{V}_{h}\times \left( 
\mathbf{\nabla }\times \mathbf{P}_{h}\right) -\mathbf{\nabla }\psi _{h},
\label{mh}
\end{eqnarray}%
where $\mathbf{P}_{e}=\mathbf{V}_{e}-\mathbf{A,}$ $\mathbf{P}_{l}=\mathbf{V}%
_{l}+\mu _{l}\mathbf{A,}$ $\mathbf{P}_{h}=\mathbf{V}_{h}+\mu _{h}\mathbf{A,}$
$\psi _{e}=-\phi +p_{e}+V_{e}^{2}/2,$ $\psi _{l}=\mu _{l}e\phi /m_{l}+\mu
_{l}p_{l}/N_{l}+V_{l}^{2}/2$ and $\psi _{h}=\mu _{h}e\phi /m_{h}+\mu
_{h}p_{h}/N_{h}+V_{h}^{2}/2$. In above equations, velocity $\mathbf{V}%
_{\alpha }\left( \alpha =e,l,h\right) \ $of the plasma species is normalized
to Alfv\'{e}n velocity $V_{A}=B_{0}/\sqrt{4\pi n_{e}e^{2}}\ $where $B_{0}$
is a constant magnetic field.$\ m_{e},\ m_{l},$ and $m_{h}$ are the masses
of electron, lighter ion and heavier ion respectively. We have defined $\mu
_{h}=Z_{h}m_{e}/m_{h}$ and $\mu _{l}=Z_{l}m_{e}/m_{l}.$ The pressure of the
plasma species normalized by $B_{0}^{2}/4\pi $ is denoted by$\ p_{\alpha }.$
The scalar potential $\phi $ is normalized to $V_{A}B_{0}\lambda _{e}$ and
vector potential $\mathbf{A\ }$is normalized to $\lambda _{e}B.$ The
magnetic field $\mathbf{B}$ is normalized by $B_{0}.$ All the distances are
normalized to electron skin depth $\lambda _{e}=\sqrt{m_{e}c^{2}/4\pi
n_{e}e^{2}}.\ $Normalization factor for time is $V_{A}/\lambda _{e}.\ $In
evaluating the above equations, we have used the relation \ $\mathbf{%
E=-\nabla }\phi -c^{-1}\partial \mathbf{A/}\partial t$, where $\phi $ is the
electrostatic potential and $c$ is the speed of light. The Eqs. (\ref{me}-%
\ref{mh}) after taking the curl read as%
\begin{eqnarray}
\frac{\partial }{\partial t}\left( \nabla \times \mathbf{P}_{e}\right)
&=&\nabla \times \left[ \mathbf{V}_{e}\times \left( \mathbf{\nabla }\times 
\mathbf{P}_{e}\right) \right] ,  \label{ce} \\
\frac{\partial }{\partial t}\left( \nabla \times \mathbf{P}_{l}\right)
&=&\nabla \times \left[ \mathbf{V}_{l}\times \left( \mathbf{\nabla }\times 
\mathbf{P}_{l}\right) \right] ,  \label{cl} \\
\frac{\partial }{\partial t}\left( \nabla \times \mathbf{P}_{h}\right)
&=&\nabla \times \left[ \mathbf{V}_{h}\times \left( \mathbf{\nabla }\times 
\mathbf{P}_{h}\right) \right] .  \label{ch}
\end{eqnarray}%
The Eqs. (\ref{ce}-\ref{ch}) can be rewritten as%
\begin{equation}
\frac{\partial \mathbf{\Omega }_{\alpha }}{\partial t}-\nabla \times \left( 
\mathbf{V}_{\alpha }\times \Omega _{\alpha }\right) =0,  \label{gve}
\end{equation}%
where $\mathbf{\Omega }_{\alpha }=\nabla \times \mathbf{P}_{\alpha }\ $are
the generalized vorticities of $\alpha $ species. The simplest steady state
solution of Eq. (\ref{gve}) can be expressed as $\mathbf{\Omega }_{\alpha
}=b_{\alpha }\mathbf{V}_{\alpha },$ where $b_{\alpha }$ $(\alpha =e,l,h)$
are arbitrary constants known as Beltrami parameters. Hence, the Beltrami
conditions for the electrons, lighter and heavier ion fluids are as follows%
\begin{eqnarray}
\nabla \times \mathbf{P}_{e} &=&b_{e}\mathbf{V}_{e},  \label{2} \\
\nabla \times \mathbf{P}_{l} &=&b_{l}\mathbf{V}_{l},  \label{3} \\
\nabla \times \mathbf{P}_{h} &=&b_{h}\mathbf{V}_{h}.  \label{4}
\end{eqnarray}%
The Eqs. (\ref{2}-\ref{4}) depict that canonical vorticities of all the
constituent species of the system follow their\ velocities. The expressions
of the relaxed equilibrium states can be derived with the help of Beltrami
conditions and the Ampere's Law

\begin{equation}
\nabla \times \mathbf{B}=N_{h}\mathbf{V}_{h}+N_{l}\mathbf{V}_{l}-\mathbf{V}%
_{e}.  \label{1}
\end{equation}%
From Eqs. (\ref{2}) and (\ref{1}), we have%
\begin{equation}
\mathbf{V}_{l}=d_{1}\left[ (\mathbf{\nabla \times })^{2}\mathbf{B}-b_{e}%
\mathbf{\nabla \times B}+d_{2}\mathbf{B}+d_{3}\mathbf{V}_{h}\right] .
\label{vl}
\end{equation}%
where $d_{1}=1/Z_{l}N_{l}\left( b_{l}-b_{e}\right) $, $d_{2}=1+\mu
_{l}Z_{l}N_{l}+\mu _{h}Z_{h}N_{h}$, and $d_{3}=\left( b_{e}-b_{h}\right)
Z_{h}N_{h}$. Putting the value of $\mathbf{V}_{l}$ into Eq. (\ref{3}), we get%
\begin{equation}
\mathbf{V}_{h}=\frac{1}{\alpha _{3}}\left[ (\mathbf{\nabla \times })^{3}%
\mathbf{B}-\left( b_{e}+b_{l}\right) (\mathbf{\nabla \times })^{2}\mathbf{B}%
+\alpha _{1}\mathbf{\nabla \times B}-\alpha _{2}\mathbf{B}\right] ,
\label{vh}
\end{equation}%
where $\alpha _{1}=1+\mu _{l}Z_{l}N_{l}+\mu _{h}Z_{h}N_{h}+b_{e}b_{l},$ $%
\alpha _{2}=b_{l}\left( 1+\mu _{l}Z_{l}N_{l}+\mu _{h}Z_{h}N_{h}\right)
+\left( b_{e}-b_{h}\right) \mu _{h}Z_{h}N_{h}-\left( b_{l}-b_{e}\right) \mu
_{l}Z_{l}N_{l}$ and $\alpha _{3}=\left( b_{e}-b_{h}\right) \left(
b_{l}-b_{h}\right) Z_{h}N_{h}.$ Eliminating $\mathbf{V}_{h}$ from Eqs. (\ref%
{4}) and (\ref{vh}), we obtain the following 4th order differential equation
in $\mathbf{B}$%
\begin{equation}
(\mathbf{\nabla \times })^{4}\mathbf{B}-a_{3}(\mathbf{\nabla \times })^{3}%
\mathbf{B}+a_{2}(\mathbf{\nabla \times })^{2}\mathbf{B}-a_{1}\mathbf{\nabla
\times B}+a_{0}\mathbf{B}=0,  \label{QBF}
\end{equation}%
where $a_{3}=b_{e}+b_{l}+b_{h},a_{2}=\alpha
_{1}+b_{e}b_{h}+b_{l}b_{h},a_{1}=b_{h}\alpha _{1}+\alpha _{2}$ and $%
a_{0}=b_{h}\alpha _{2}+\mu _{h}\alpha _{3}$. Equation (\ref{QBF}) is
equivalent to the sum of four Beltrami fields and it is called Quadruple
Beltrami (QB) state. This is the equilibrium solution of the system provided
the Bernoulli conditions $\mathbf{\nabla }\psi _{\alpha }=0$ are satisfied.
The Bernoulli conditions are important but in the present work, they are not
directly relevant. Substituting the value of $\mathbf{V}_{h}$ into Eq. (\ref%
{vl}), we get%
\begin{equation}
\mathbf{V}_{l}=l_{4}\mathbf{\nabla \times \nabla \times \nabla \times B}%
-l_{3}\mathbf{\nabla \times \nabla \times B}+l_{2}\mathbf{\nabla \times B}%
-l_{1}\mathbf{B},  \label{vlb}
\end{equation}%
where $l_{4}=\alpha _{6}k_{4}/\alpha _{4},$ $l_{3}=(\alpha
_{6}k_{3}-1)/\alpha _{4},$ $l_{2}=(\alpha _{6}k_{2}-b_{e})/\alpha _{4},$ $%
l_{1}=(\alpha _{6}k_{1}-\alpha _{4})/\alpha _{4},$ $\alpha
_{4}=Z_{l}N_{l}\left( b_{l}-b_{e}\right) ,\mathbf{\ }\alpha _{5}=1+\mu
_{l}Z_{l}N_{l}+\mu _{h}Z_{h}N_{h}$\textbf{, }and\textbf{\ }$\alpha
_{6}=\left( b_{e}-b_{h}\right) Z_{h}N_{h}.$ Substituting these values of $%
\mathbf{V}_{h}$ and $\mathbf{V}_{l}$ in Eq. (\ref{1}), we obtain%
\begin{equation}
\mathbf{V}_{e}=e_{4}\mathbf{\nabla \times \nabla \times \nabla \times B}%
-e_{3}\mathbf{\nabla \times \nabla \times B}+e_{2}\mathbf{\nabla \times B}%
-e_{1}\mathbf{B,}  \label{veb}
\end{equation}%
where $e_{4}=Z_{l}N_{l}l_{4}+Z_{h}N_{h}k_{4}$, $%
e_{3}=Z_{l}N_{l}l_{3}+Z_{h}N_{h}k_{3}$, $%
e_{2}=Z_{l}N_{l}l_{2}+Z_{h}N_{h}k_{2}-1$, and $%
e_{1}=Z_{l}N_{l}l_{1}+Z_{h}N_{h}k_{1}$. The composite velocity is given by%
\begin{equation}
\mathbf{V}=\frac{\rho _{e}\mathbf{V}_{e}+\rho _{l}\mathbf{V}_{l}+\rho _{h}%
\mathbf{V}_{h}}{\rho _{e}+\rho _{l}+\rho _{h}},  \label{VB}
\end{equation}%
where $\rho =\rho _{e}+\rho _{l}+\rho _{h}$ is the\ total fluid density, $%
\rho _{e}=m_{e}n_{e},$ $\rho _{l}=m_{l}Z_{l}n_{l,}$ and $\rho
_{h}=m_{h}Z_{h}n_{h,}$ are fluid densities of electrons, lighter ions and
heavier ions species, respectively. By simplifying Eq. (\ref{VB}), we have%
\begin{equation}
\mathbf{V}=p_{4}\mathbf{\nabla \times \nabla \times \nabla \times B}-p_{3}%
\mathbf{\nabla \times \nabla \times B}+p_{2}\mathbf{\nabla \times B}-p_{1}%
\mathbf{B,}  \label{composite}
\end{equation}%
where $p_{4}=(e_{4}+\alpha _{l}l_{4}+\alpha _{h}h_{4})/\alpha _{7},\
p_{3}=(e_{3}+\alpha _{l}l_{3}+\alpha _{h}h_{3})/\alpha _{7},\
p_{2}=(e_{2}+\alpha _{l}l_{2}+\alpha _{h}h_{2})/\alpha _{7},$ $%
p_{1}=(e_{1}+\alpha _{l}l_{1}+\alpha _{h}h_{1})/\alpha _{7},$ $\alpha
_{l}=Z_{l}N_{l}m_{l}/m_{e}$, $\alpha _{h}=Z_{h}N_{h}m_{h}/m_{e}$, $\alpha
_{7}=(m_{e}+Z_{l}N_{l}m_{l}+Z_{h}N_{h}m_{h})/m_{e}$, $h_{4}=1/\alpha _{3},$ $%
h_{3}=(b_{e}+b_{l})/\alpha _{3},$ $h_{2}=\alpha _{1}/\alpha _{3},$ and $%
h_{1}=\alpha _{2}/\alpha _{3}$.

To find the solution of Eq. (\ref{QBF}), we re-write it as%
\begin{equation}
(\mathbf{\nabla \times }-\lambda _{1})(\mathbf{\nabla \times }-\lambda
_{2})\left( \mathbf{\nabla \times }-\lambda _{3}\right) \left( \mathbf{%
\nabla \times }-\lambda _{4}\right) \mathbf{B}=0,  \label{QBF1}
\end{equation}%
and take%
\begin{eqnarray}
a_{0} &=&\lambda _{1}\lambda _{2}\lambda _{3}\lambda _{4},  \label{a0} \\
a_{1} &=&\lambda _{1}\lambda _{2}\left( \lambda _{3}+\lambda _{4}\right)
+\lambda _{3}\left( \lambda _{1}\lambda _{4}+\lambda _{2}\lambda _{4}\right)
,  \label{a1} \\
a_{2} &=&\lambda _{1}\left( \lambda _{2}+\lambda _{3}+\lambda _{4}\right)
+\lambda _{2}\left( \lambda _{3}+\lambda _{4}\right) +\lambda _{3}\lambda
_{4},  \label{a2} \\
a_{3} &=&\lambda _{1}+\lambda _{2}+\lambda _{3}+\lambda _{4}.  \label{a3}
\end{eqnarray}%
Writing the curl operator \textquotedblleft $\mathbf{\nabla \times }$%
\textquotedblright\ as curl, the equation (\ref{QBF1}) can be expressed as 
\begin{equation}
(curl-\lambda _{1})(curl-\lambda _{2})(curl-\lambda _{3}) (curl-\lambda _{4}) \mathbf{B}=0.
\label{curl}
\end{equation}%
The general solution of Eq. (\ref{curl}) can be written as a superposition
of four linear Beltrami fields as given below%
\begin{equation}
\mathbf{B}=C_{1}\mathbf{B}_{1}+C_{2}\mathbf{B}_{2}+C_{3}\mathbf{B}_{3}+C_{4}%
\mathbf{B}_{4}.
\end{equation}%
The Beltrami fields $\mathbf{B}_{j}$ $\left( j=1,2,3,4\right) $ satisfy the
Beltrami condition $\mathbf{\nabla }\times \mathbf{B}_{j}=\lambda _{j}%
\mathbf{B}_{j},$ and $\lambda _{j}$ $\left( j=1,2,3,4\right) $ are\ the
eigenvalues of curl operator. Dimensions of the curl operators are the
inverse of length and called scale parameters. From equations (\ref{a0}-\ref%
{a3}), we find that the scale parameters are the roots of\ the following\
quartic equation%
\begin{equation}
\lambda ^{4}-a_{3}\lambda ^{3}+a_{2}\lambda ^{2}-a_{1}\lambda +a_{0}=0.
\label{quartic}
\end{equation}%
The four roots of Eq. (\ref{quartic}) are%
\begin{eqnarray}
\lambda _{1} &=&\frac{a_{3}+2S+2\gamma }{4},  \label{l1} \\
\lambda _{2} &=&\frac{a_{3}+2S-2\gamma }{4},  \label{l2} \\
\lambda _{3} &=&\frac{a_{3}-2S+2\xi }{4},  \label{l3} \\
\lambda _{4} &=&\frac{a_{3}-2S-2\xi }{4},  \label{l4}
\end{eqnarray}%
where $S=\left( \sqrt{\beta _{3}^{2}-4\beta _{2}+4Y}\right) /2,$ $Y=\left(
d-3u^{2}+3u\beta _{1}\right) /3u,$ $d=\left( 3\beta _{2}-\beta
_{1}^{2}\right) /3,$ $u=\sqrt[3]{\left( q/2\right) +\left( q/2\right)
^{2}+\left( d/3\right) ^{3}},$ $q=\left( 9\beta _{1}\beta _{2}-2\beta
_{1}-27\beta _{3}\right) /27,$ $\beta _{1}=a_{2},$ $\beta
_{2}=a_{1}a_{3}-4a_{0},$ $\beta _{3}=a_{1}^{2}+a_{3}^{2}a_{0}-4a_{2}a_{0},$ $%
\gamma =\sqrt{\frac{3}{4}a_{3}^{2}-S^{2}-2a_{2}+\frac{1}{4S}%
(4a_{3}a_{2}-8a_{1}-a_{3}^{3})},$ and $\xi =\sqrt{\frac{3}{4}%
a_{3}^{2}-S^{2}-2a_{2}-\frac{1}{4S}(4a_{3}a_{2}-8a_{1}-a_{3}^{3})}.$

\section{Impact of Beltrami Parameters on relaxed states}

The Beltrami parameters $\left( b_{e},b_{l},b_{h}\right) $ are the ratios of
the generalized vorticities to the fluid velocities. These parameters play
an essential role in the formation of relaxed states. Let us now describe
how the relaxed states are affected by them.

\begin{description}
\item[(a)] When all the Beltrami parameters are equal to each other ($%
b_{e}=b_{l}=b_{h}=b$),\ then the steady state equilibrium satisfies the DB
equation \cite{Mahajan1998}.%
\begin{equation}
\left( \mathbf{\nabla \times }\right) ^{2}\mathbf{B}-a_{1}\mathbf{\nabla
\times B}+a_{0}\mathbf{B}=0,  \label{db}
\end{equation}%
where $a_{1}=b,$ and $a_{0}=1+Z_{h}N_{h}\mu _{h}+Z_{l}N_{l}\mu _{l}$. It
shows that when the ratios of the canonical vorticities to the velocities of
all the plasma fluids are equal, the steady state equilibria permit only two
distinct Beltrami fields.

\item[(b)] When the generalized vorticities of all the plasma components
vanish such that $b_{e}=b_{l}=b_{h}=0,$ then%
\begin{equation}
\mathbf{\nabla \times \nabla \times B}+a_{1}\mathbf{B=}0,  \label{le}
\end{equation}%
where $a_{1}=1+Z_{l}N_{l}\mu _{l}+Z_{h}N_{h}\mu _{h}.$ The above equation (%
\ref{le}) is the well known London's equation and it is the manifestation of
classical perfect diamagnetic state. The relaxed equilibria is governed by
two Beltrami fields \cite{Mahajan2008}.

\item[(c)] When the ratio of the generalized vorticities to the
corresponding velocities of lighter and heavier ions are equal ($%
b_{l}=b_{h}=b$),\ the system will be self-organized to Triple Beltrami (TB)
state \cite{Iqbal2012},%
\begin{equation}
\left( \mathbf{\nabla \times }\right) ^{3}\mathbf{B}-a_{2}\left( \mathbf{%
\nabla \times }\right) ^{2}\mathbf{B}+a_{1}\mathbf{\nabla \times B}+a_{0}%
\mathbf{B}=0,  \label{tcb}
\end{equation}%
where $a_{2}=b_{e}+b,\ a_{1}=1+Z_{h}N_{h}\mu _{h}+Z_{l}N_{l}\mu _{l}+b_{e}b$
and $a_{0}=\left( Z_{h}N_{h}\mu _{h}+Z_{l}N_{l}\mu _{l}\right) \left(
b-b_{e}\right) -b(1+Z_{h}N_{h}\mu _{h}+Z_{l}N_{l}\mu _{l}).$ In this case,
the system allow three Beltrami fields on relaxation.

\item[(d)] When \ two Beltrami parameters are set equal to zero (let $%
b_{e}=b_{l}=0$), then the relaxed state is given by TB equation%
\begin{equation}
\left( \mathbf{\nabla \times }\right) ^{3}\mathbf{B}-a_{2}\left( \mathbf{%
\nabla \times }\right) ^{2}\mathbf{B}+a_{1}\mathbf{\nabla \times B}+a_{0}%
\mathbf{B}=0,  \label{tcb1}
\end{equation}%
where $a_{2}=b_{h},$ $a_{1}=1+Z_{l}N_{l}\mu _{l}+Z_{h}N_{h}\mu _{h}$ and $%
a_{0}=b_{h}\left( Z_{l}N_{l}\mu _{l}+1\right) $ and three Beltrami states
exist at the equilibrium.

\item[(e)] When the vorticity of any one of the plasma species is set equal
to zero, say $b_{e}=0$, we obtain QB\ relaxed state with modified
coefficients as given below%
\begin{equation}
(\mathbf{\nabla \times )}^{4}\mathbf{B-}a_{3}(\mathbf{\nabla \times )}^{3}%
\mathbf{B}+a_{2}\left( \mathbf{\nabla \times }\right) ^{2}\mathbf{B}-a_{1}%
\mathbf{\nabla \times B}+a_{0}\mathbf{B}=0,  \label{qcb}
\end{equation}%
where $a_{3}=b_{l}+b_{h},$ $a_{2}=$ $1+Z_{h}N_{h}\mu _{h}+Z_{l}N_{l}\mu
_{l}+b_{l}b_{h},a_{1}=b_{l}\left( 1+Z_{h}N_{h}\mu _{h}\right)
-b_{h}Z_{h}N_{h}\mu _{h}+\left( 1+Z_{h}N_{h}\mu _{h}+Z_{l}N_{l}\mu
_{l}\right) $ and $a_{0}=b_{l}b_{h}(1+Z_{h}N_{h}\mu _{h}+Z_{l}N_{l}\mu
_{l})-b_{l}Z_{l}N_{l}\mu _{l}-b_{h}Z_{h}N_{h}\mu _{h}.$ This state has
maximum number of Beltrami fields equal to four.
\end{description}

\section{Quadruple Beltrami Magnetic Fields}

The magnetic fields and velocities permitted by the system (see equations (%
\ref{QBF}),(\ref{composite}),(\ref{QBF1})) have four characteristic length
scales $\left( \lambda _{j}^{-1}\right) $. The length scales are determined
by the amounts of generalized helicities present in the system. These scales
are functions of the Beltrami parameters $\left( b_{e},b_{l},b_{h}\right) ,$
masses and number densities of the plasma species. In order to see the
impact of four inherent scales, we need to study the explicit solutions of
equation (\ref{QBF1}). The explicit solutions of equation (\ref{QBF1}) are
the three dimensional ABC flow in Cartesian coordinates \cite%
{Arnold1998,YoshidaBook} and Chandrasekhar \& Kendall functions in
cylindrical coordinates $(r,\theta ,z)$ \cite{Kendall1957}. Considering
plasma in a cube of length \textit{L,} a two dimensional \textit{ABC} field
in Cartesian coordinates have been used in modeling the coronal structures
evolved by DB states and to derive the conditions for catastrophic
transformations of states and conversion of magnetic energy to the flow
kinetic energy \cite{Ohsaki2001,Ohsaki2002}. The Kelvin-Helmholtz
instability in Beltrami fields has been investigated using slab geometry in
Cartesian coordinates $(x,y,z)$ \cite{Ito2002}. In this present work, we
also use the slab geometry to illustrate the influence of four
characteristic scale lengths on the formation of self-organized magnetic
fields and velocities. We assume that the QB magnetic fields and velocities
are functions of $x$ only and they have only $y$ and $z$ components. We
consider the plasma in the region $-x_{0}\leq x\geq x_{0}$ with the center
at $x=0$. In the slab geometry, the QB magnetic field can be written as%
\begin{equation}
\mathbf{B}=\sum\limits_{j=1}^{4}C_{j}\left( 
\begin{array}{c}
0 \\ 
\sin \lambda _{j}x \\ 
\cos \lambda _{j}x%
\end{array}%
\right) ,\,\left\vert x\right\vert \leq x_{0}  \label{slab}
\end{equation}%
The value of constants $C_{j}$ can be calculated with the help of boundary
conditions. By applying the boundary conditions $\left\vert B_{y}\right\vert
_{x=x_{0}}=g,$ $\left\vert B_{z}\right\vert _{x=0}=h,$ $\left\vert (\mathbf{%
\nabla }\times \mathbf{B)}_{z}\right\vert _{x=x_{0}}=w,$ and $\left\vert (%
\mathbf{\nabla }\times \mathbf{B)}_{y}\right\vert _{x=0}=u$, the constants
are given by%
\begin{equation}
C_{j}=\frac{\zeta _{j}}{\kappa },\text{ }j=1,2,3,4.
\end{equation}%
where%
\begin{align}
\zeta _{1}& =[T_{2}\sin (\lambda _{3}x_{0})\sin (\lambda
_{4}x_{0})+W_{2}\sin (\lambda _{2}x_{0})](\lambda _{3}-\lambda _{4}) \\
& +[T_{4}\sin (\lambda _{3}x_{0})\sin (\lambda _{2}x_{0})+W_{4}\sin \lambda
_{4}x_{0})](\lambda _{2}-\lambda _{3})  \notag \\
& +[T_{3}\sin (\lambda _{2}x_{0})\sin (\lambda _{4}x_{0})+W_{3}\sin (\lambda
_{3}x_{0})](\lambda _{4}-\lambda _{2}),  \notag
\end{align}%
\begin{align}
\zeta _{2}& =[T_{4}\sin (\lambda _{3}x_{0})\sin (\lambda
_{1}x_{0})+W_{4}\sin (\lambda _{4}x_{0})](\lambda _{3}-\lambda _{1}) \\
& +[T_{1}\sin (\lambda _{3}x_{0})\sin (\lambda _{4}x_{0})+W_{1}\sin (\lambda
_{1}x_{0})](\lambda _{4}-\lambda _{3})  \notag \\
& +[T_{3}\sin (\lambda _{1}x_{0})\sin (\lambda _{4}x_{0})+W_{3}\sin (\lambda
_{3}x_{0})](\lambda _{1}-\lambda _{4}),  \notag
\end{align}%
\begin{align}
\zeta _{3}& =[T_{2}\sin (\lambda _{1}x_{0})\sin (\lambda
_{4}x_{0})+W_{2}\sin (\lambda _{2}x_{0})](\lambda _{4}-\lambda _{1}) \\
& +[T_{4}\sin (\lambda _{1}x_{0})\sin (\lambda _{2}x_{0})+W_{4}\sin (\lambda
_{4}x_{0})](\lambda _{1}-\lambda _{2})  \notag \\
& +[T_{1}\sin (\lambda _{2}x_{0})\sin (\lambda _{4}x_{0})+W_{1}\sin (\lambda
_{1}x_{0})](\lambda _{2}-\lambda _{4}),  \notag
\end{align}%
\begin{align}
\zeta _{4}& =[T_{2}\sin (\lambda _{3}x_{0})\sin (\lambda
_{1}x_{0})+W_{2}\sin (\lambda _{2}x_{0})](\lambda _{1}-\lambda _{3}) \\
& +[T_{3}\sin (\lambda _{1}x_{0})\sin (\lambda _{2}x_{0})+W_{3}\sin (\lambda
_{3}x_{0})](\lambda _{2}-\lambda _{1})  \notag \\
& +[T_{1}\sin (\lambda _{2}x_{0})\sin (\lambda _{3}x_{0})+W_{1}\sin (\lambda
_{1}x_{0})](\lambda _{3}-\lambda _{2}),  \notag
\end{align}%
and%
\begin{align}
\kappa & =[\sin (\lambda _{1}x_{0})\sin (\lambda _{4}x_{0})+\sin (\lambda
_{2}x_{0})\sin (\lambda _{3}x_{0})]K_{1} \\
& +[\sin (\lambda _{3}x_{0})\sin (\lambda _{4}x_{0})+\sin (\lambda
_{2}x_{0})\sin (\lambda _{1}x_{0})]K_{2}  \notag \\
& +[\sin (\lambda _{1}x_{0})\sin (\lambda _{3}x_{0})+\sin (\lambda
_{2}x_{0})\sin (\lambda _{4}x_{0})]K_{3},  \notag
\end{align}%
where $u-h\lambda _{j}=T_{j},$ and $\ w-g\lambda _{j}=W_{j}$ with $%
j=1,2,3,4. $ $K_{1}=(\lambda _{3}-\lambda _{2})(\lambda _{4}-\lambda _{1}),$ 
$K_{2}=(\lambda _{2}-\lambda _{1})(\lambda _{4}-\lambda _{3})$ and $%
K_{3}=(\lambda _{3}-\lambda _{1})(\lambda _{2}-\lambda _{4}).$ The composite
velocity is given by%
\begin{equation}
\mathbf{V}=\sum\limits_{j=1}^{4}C_{j}s_{j}%
\begin{pmatrix}
0 \\ 
\sin \lambda _{j}x \\ 
\cos \lambda _{j}x%
\end{pmatrix}%
,\,\left\vert x\right\vert \leq x_{0}
\end{equation}%
where $s_{1}=\lambda _{1}^{3}p_{4}-\lambda _{1}^{2}p_{3}+\lambda
_{1}p_{2}-p_{1},$ $s_{2}=\lambda _{2}^{3}p_{4}-\lambda _{2}^{2}p_{3}+\lambda
_{2}p_{2}-p_{1},$ $s_{3}=\lambda _{3}^{3}p_{4}-\lambda _{3}^{2}p_{3}+\lambda
_{3}p_{2}-p_{1},$ and $s_{4}=\lambda _{4}^{3}p_{4}-\lambda
_{4}^{2}p_{3}+\lambda _{4}p_{2}p_{1}$.

The Quadruple Beltrami magnetic field is characterized by four spatial scale
parameters which are functions of Beltrami parameters, inertia, number
density and amount of charges carried by two positive ions. Hence, the
system allows a wide variety of multiscale self-organized states ranging
from force-free paramagnetic to fully diamagnetic states that result due to
interaction of plasma flows with the magnetic field as shown in the Ref. 
\cite{Gondal2017}. However, the present solutions in contrast to the
previous work shown in \cite{Gondal2017}, allow us to create and study the
multiscale structures depending on the inertia and charges of the two
positive ions. Hence, we can apply the present work to analyze the dynamics
and characteristics of the magnetic field structures and the large scale
velocity fields observed in plasmas composed of two positive ions and
electrons which are found in different laboratory and astrophysical
environments, for instance, the plasmas found in the Earth's
magnetotail/magnetosheath, at the bow shock in front of the magnetopause
boundary layer \cite{stasiewicz}, in the heliosphere \cite{schwenn} and in
the magnetic equator region of the Saturn \cite{Shemansky1992}.

\section{Scale separation and its implications}

There exist certain physical phenomena in which processes at microscopic
scales are thought to be responsible for the evolution of macroscopic
structures e.g., reconnection of flux tubes, flares and accretion disks etc.
It is therefore logical to deduce and understand the observed macroscopic
phenomena on the basis of microdynamics. In multi fluid plasmas, the Hall
effect introduces multi scales in contrast to single fluid
magnetohydrodynamics (MHD) plasmas. For instance, Hall MHD plasmas allow two
disparate scales which interact each other. One of the scales is microscopic
and corresponds to system size while the other scale is the intrinsic scale
(ion skin depth) of Hall MHD at which kinetic effects of the ions become
important. The Hall term introduces the singular perturbation which connects
the two disparate scales and leads to two interacting self-organized states 
\cite{ZMO2004,OZ2005}

The singular perturbation introduced by Hall effect causes a strong mutual
interaction between the magnetic field and velocity in multi-fluid plasmas.
This interaction results in the appearance of four structures corresponding
to four scale parameters ($\lambda _{j};$ $j=1,2,3,4)$ when the multi-ion
plasma comprising of three components attains the steady state equilibrium.
We note that the scale parameters are the eigenvalues of the curl operator
and characterize the reciprocal of the length scales on which both the
magnetic fields and velocity fields change significantly. The scale
parameters can assume quite different magnitudes on varying Beltrami
parameters, inertia and densities of the species. Hence, the scale
parameters may be distinct having very different values. At least one of the
scale parameters may be very small indicating a very large structure
(macroscopic structure) of the order of the system size and one of them may
be very large showing the existence of very small structure (microscopic
structure) of the order of the characteristic length of the system. The
other scale parameters, if they are different, may be of intermediate sizes
or less than the ion skin depth \cite{Lingam2015}.

This scale separation gives rise to very interesting phenomena. The magnetic
field may be regular and smooth and can couple with the velocity which may
have jittery or oscillatory profile. It happens when the magnetic field is
dominated by the long scale field $\left( \left\vert \lambda _{j}\right\vert
\ll 1\right) ,$ while the velocity field may have a large component varying
on the short scale $\left( \left\vert \lambda _{j}\right\vert \gg 1\right) $%
. The opposite case may also pertain.

\begin{figure}[h]
\centering
\includegraphics{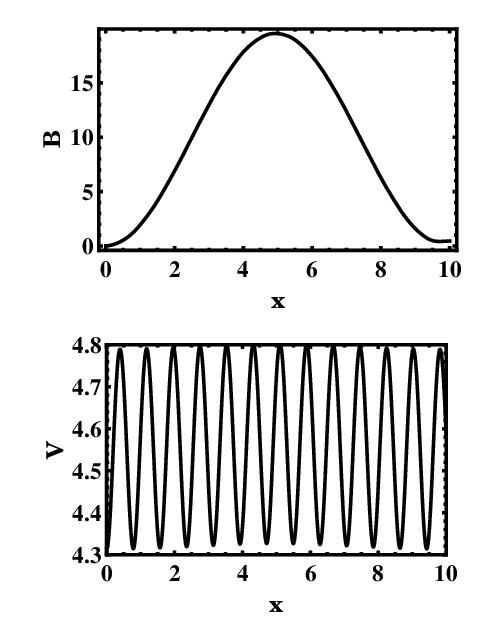}
\caption{Profiles of magnetic field and velocity for Beltrami parameters $b_{e}=2.1$, $b_{l}=2.0$ and $b_{h}=10.0.$}
\label{sb}
\end{figure}

\begin{figure}[h]
\centering
\includegraphics{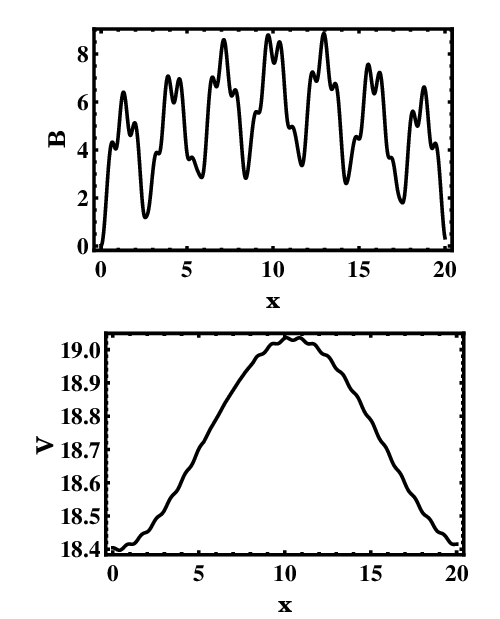}
\caption{Profiles of magnetic
field and velocity for Beltrami parameters $b_{e}=9.9,$ $b_{l}=2.0$ and $
b_{h}=2.3.$}
\label{sv}
\end{figure}

In order to show the scale separation of two fields, we consider the
multi-ion plasmas containing electrons and significant amounts of two
species of positively charged ions. Though such plasmas are found abundantly
in space and laboratory as mentioned in the introduction, however, we
consider the e-H$^{\text{+}}$-O$^{\text{+}}$ plasma found in the magnetic
equator region of the Saturn for the present analysis \cite{Shemansky1992}.
We take $n_{H^{+}}=n_{O^{+}}=0.5n_{e}$, where $n_{e},$ $n_{H^{+}}$ and $%
n_{O^{+}}$ are densities of electrons, hydrogen ions and oxygen ions,
respectively. To plot the graphs shown in figures (\ref{sb}) and (\ref{sv}),
the boundary conditions are taken as $u=0.3,$ $h=0.25,$ $w=0.02,$ and $%
g=0.017.$ The electron skin depth for this plasma is $\sim 9.3\times 10^{4}$
cm. Let us take two sets of Beltrami parameters to achieve the scale
separation. First, we take the set of Beltrami parameters $b_{e}=2.1,$ $%
b_{l}=2.0,$ and $b_{h}=10.0$. The corresponding scale parameters are $%
\lambda _{1}=0.73,\ \lambda _{2}=1.37,\ \lambda _{3}=2.0,\ $and $\lambda
_{4}=10$. The scales of the structures are given by $l_{j}=\lambda
_{e}/\lambda _{j}.$ Hence, in this case, the system exhibits four structures
of lengths $l_{1}=1.3\times 10^{5}$ cm$,$ $l_{2}=6.8\times 10^{4}$ cm$,$ $%
l_{3}=4.7\times 10^{4}$ cm$,$ and $l_{4}=9.3\times 10^{3}$ cm, respectively.
The interaction among these structures gives rise to a smooth magnetic field
coupled with a jittery velocity as shown in figure (\ref{sb}). This
situation is of fundamental significance as it exposes the mechanism
operating in the systems where one finds smooth magnetic fields in
conjunction with the irregular flows. One of the examples is the generic
turbulent dynamo. The turbulent dynamo is the process where the complex
flows generate the ordered magnetic fields. This is the scale separation
process that shows the possibility of the existence of small-scale velocity
field energy reservoir which can be transformed to smooth magnetic field
(amplifying it) while these results can be obtained by time-dependent dynamo
process only under specific conditions \cite%
{Mininni2002,Mininni2003,Lingam2015,Mininni2007}.

Another important implication of the scale separation is the heating of
plasma through a dissipative process due to the emergence of short-scale
velocity fields. The short-scale velocities could generate an appreciable
amount of large viscous dissipation. It is possible when the magnetic field
and flows have different length scales. In the presence of a smooth magnetic
field if one assumes that both the magnetic and velocity fields have same
length scales, then a reasonable dissipation of energy can not be produced
in the system either by resistive or viscous damping and hence the plasma
can not be heated. However, the present results show that smooth magnetic
fields (for instance, those produced by turbulent dynamo) can be associated
with big and jittery flows which can significantly heat the multi-ions
plasma when it is damped by a finite viscosity. Although, the present theory
does not contain viscosity, it suggests an important mechanism where the
magnetic fields and flows with far different length scales couple each
other. A possible consequence could be the heating of the plasma by the
viscous dissipation of the jittery part of the flow kinetic energy even for
relatively smooth magnetic fields as it has been shown by Mahajan et. al. 
\cite{Mahajan2001} that viscous damping associated with the fastly varying
(spatially) velocity field causes the primary heating needed to create and
maintain the solar corona bright.

Next, we suppose $b_{e}=9.9,$ $b_{l}=2.0,$ and $b_{h}=2.3$ with the
corresponding scale parameters $\lambda _{1}=0.10,\ \lambda _{2}=2.0,\
\lambda _{3}=2.3,\ $and $\lambda _{4}=9.8.$ The sizes of the structures are
given by $l_{1}=9.3\times 10^{5}$ cm$,$ $l_{2}=4.7\times 10^{4}$ cm$,$ $%
l_{3}=4.0\times 10^{4}$ cm$,$ and $l_{4}=9.5\times 10^{3}$ cm, respectively.
In contrast to the magnetic field and flow profiles shown in figure (\ref{sb}%
), we find that a short scale magnetic field now couples with a smooth flow
as shown in figure (\ref{sv}). Again this is the scale separation that shows
the possibility of the existence of small-scale magnetic field energy
reservoir which can be transformed to smooth velocity field (amplifying it)
while the similar field configurations can be obtained under specific
conditions only by using time-dependent inverse dynamo process \cite%
{Mahajanshata,Lingam2015,Mahajan2005a,Mininni2007,Kotorashvili2020,Kotorashvili2022}%
.

\section{Summary}

The formation of multiscale self-organized structures in multi-ion plasma
whose constituents are electrons, lighter and heavier ions has been
investigated. The steady state solutions of vorticity evolution equations
satisfy the Beltrami conditions which show strong coupling of velocities and
magnetic fields. On solving the Beltrami conditions along with the Ampere's
law, the expression for the relaxed magnetic field is evaluated which
satisfy the QB equation. The QB state have a Beltrami index of four and
represents the superposition of four Beltrami fields. The eigenvalues of the
four Beltrami fields are the roots of a quartic equation. The eigenvalues
(scale parameters) being inverse of length represent the length scales on
which the fields and velocities change significantly. This disparate
variation of the magnetic field and velocity may cause heating of multi-ion
plasma through dissipation. The disparate variation of the velocity and
magnetic field is precisely the required recipe which lies at the heart of
the dynamo mechanism: the dynamo in which the short scale velocity field
produces a relatively smooth magnetic field, and the reverse dynamo in which
the length scales of the two fields are reversed. In the present work, we
have just shown that the seeds of both these possibilities are there even in
the simplest manifestation of the steady state equilibria of the multi-ion
magnetized plasmas.

The present study and results have potential relevance to a variety of
phenomena observed in laboratory, space and cosmic plasmas. A vast spectrum
of problems in multi-ion plasmas can be studied and investigated in the
context of QB fields. For instance, highly confining, fully diamagnetic and
minimum $\left\vert B\right\vert $ configurations could be constructed
following the Refs. \cite{Steinhauer1998,Mahajan1998}. The QB states could
also be used to model the catastrophic eruptions, such as solar coronal mass
ejections (CME's), supernovae and gamma ray bursts as it has been studied in
the context of DB formulation \cite{Ohsaki2001,Ohsaki2002,Mahajan2002}. It
is concluded that QB states are of considerable importance and could be
employed to study the formation of multiscale structures in a diverse range
of terrestrial and astrophysical phenomena occurring in plasmas composed of
electrons and two different groups of positive ions.

\begin{acknowledgments}
The authors are thankful to the anonymous reviewer for his suggestions and
guidance to improve the quality of the work. The work of M. Iqbal was
supported by Higher Education Commission (HEC), Pakistan under Project No.
20-9408/Punjab/NRPU/R\&D/HEC/2017-18
\end{acknowledgments}

\end{document}